\documentclass[aps,epsf,preprint]{revtex4-1}
\usepackage{amsmath,amssymb}

\def\1{{\bf 1}}
\def\[{\left[}
\def\]{\right]}
\def\be{\begin{eqnarray}}
\def\ee{\end{eqnarray}}
\def\bm{\begin{pmatrix}}
\def\em{\end{pmatrix}}
\def\nn{\nonumber}
\def\({\left(}
\def\){\right)}

\def\eq#1{Eq.(\ref{#1})}

\def\a{\alpha}
\def\r{\rho}
\def\s{\sigma}

\def\e{\epsilon}

\def\f{\phi}

\def\G{{\Gamma}}

\def\l{\lambda}

\def\x{\times}

\def\p{\partial}

\def\l{\lambda}
\def\n{\nu}
\def\h{{1\over 2}}

\def\labels#1{\label{#1}}
\def\edc{\end{document}}

\def\bn{\begin{enumerate}}
\def\i{\item}
\def\en{\end{enumerate}}
\def\b{\beta}
\def\g{\gamma}

\def\ba{\begin{array}}
\def\ea{\end{array}}
\def\bc{\begin{center}}
\def\ec{\end{center}}

\def\edoc{\end{document}}

\def\^{$\wedge$}
\def\.{\!\cdot\!}

\def\igw#1{\includegraphics[width=#1cm]}
\def\+{\!+\!}
\def\-{\!-\!}
\def\={\!=\!}
\def\vs{\vskip.5cm}

\usepackage{color}

\def\M{M\"obius\ }
\def\Q{\Psi}
\def\pf{{\rm Pf}}
\def\Tr{{\rm Tr}}
\def\igw#1{\includegraphics[width=#1cm]}
 
\usepackage{graphicx}
\graphicspath{/Users/harrylam/Current/filetex/Helicity/graphics/}

\begin{document}
\title{Pfaffian Diagrams for Gluon Tree Amplitudes}
\author{C.S. Lam$^{1,2,3}$}
\email{Lam@physics.mcgill.ca}
\address{$^1$Department of Physics, McGill University\\
 Montreal, Q.C., Canada H3A 2T8\\
$^2$Department of Physics and Astronomy, University of British Columbia,  Vancouver, BC, Canada V6T 1Z1 \\
$^3$CAS Key Laboratory of Theoretical Physics, Institute of Theoretical Physics, Chinese Academy of
Sciences, Beijing 100190, China\\}

\begin{abstract}

Pfaffian diagrams are formulated to represent gluon amplitudes computed from the Cachazo-He-Yuan (CHY) formula. 
They may be regarded
as a systematic regrouping of Feynman diagrams after internal momenta are expanded  and products
of vertex factors are evaluated. This reprocessing enables
gluon amplitudes expressed in Pfaffian diagrams  to contain  less terms. For example, there are 19 terms for the
four-point amplitude in Pfaffian diagrams, and 35 terms in Feynman diagrams.
Gauge invariance is simpler and more explicit  in Pfaffian diagrams, in that subset
of diagrams with the same root configuration are already gauge invariant in all lines but two. In getting to
these results, several technical difficulties must be overcome. Double poles must be converted to 
simple poles, integrations must be carried out directly and formulated into simple rules, and 
the three \M constant lines must be suitably chosen to minimize the number of terms present.

\end{abstract}
\narrowtext
\maketitle
\section{Introduction}
It is well known that the $n$-gluon scattering amplitude contains many terms, even in the tree approximation.
For $n=4$, there are four Feynman  diagrams, the $s, t, u$-channel  as well as the four-gluon diagrams.
Each of the former three consists of $3^2=9$ terms, and the four-gluon diagram contains 3 terms, making
a total of $3\!\x\!9\+3=30$ terms. 
To express the amplitude in measurable quantities, the polarization vectors $\e_i$
and the outgoing momenta $k_i$, internal momenta must be  expanded into sums of $k_i$,
thereby further increasing the number of terms. 
For larger $n$, the number of diagrams
grows rapidly, reaching over ten million for $n=10$. The number of terms in each diagram also increases
exponentially, being already $3^{n-2}$ if the diagram consists of triple-gluon vertices alone, and much more after
the internal momenta are expanded into sums of external momenta. 

Simplification occurs after the color factor, say ${\rm Tr}(\l_1\l_2\cdots \l_n)$, is factored out. The resulting color-stripped
amplitude has less terms, and is cyclically invariant in  the order of the color trace.
Only such
color-stripped amplitudes with the natural order  $(123\cdots n)$ will be considered in this article.
Even so, the number of terms   is still huge, so any method that
can further reduce it would be welcome. This note is an attempt to do that by formulating the diagrams in a completely
different way.

In this connection, one might think of using the Cachazo-He-Yuan (CHY) \cite{CHY1,CHY2,CHY3,CHY4,CHY5} formula, since the
tree amplitudes are given there by a {\it single}  integration over $n\-3$ variables $\s_i$. This formula, valid for 
any number of dimensions, is very efficient in understanding general properties such as gauge
invariance in gluon scattering, but to obtain explicit expression in terms of external momenta and polarizations,
the $n\-3$ integrations must still be carried out. Unfortunately, in the case of bi-adjoint $\f^3$ amplitudes,
 integration simply reverts the amplitude back to a
sum of Feynman  diagrams.

Nevertheless, there is still hope for  gluon amplitudes because the CHY formula contains no internal momenta
so at least extra terms coming from their expansion are avoided. 
Its dependence in the numerator on $\e_i$ and $k_i$ comes directly from the `reduced Pfaffian'
in the integrand. To get the gluon amplitude from the CHY formula, this reduced Pfaffian must be expanded, and the $n\-3$ integrations carried out. There are a number of technical difficulties to be overcome, but at the end 
a set of 
`Pfaffian diagrams' and `Pfaffian rules' can be devised to express the final result. These new diagrams
can be regarded as a regrouping of the familiar Feynman diagrams and Feynman rules,
although this regrouping cannot be easily derived without the CHY formalism. When the amplitude is expressed in
Feynman diagrams, a single diagram contains terms with a common denominator, the product of the propagators
in the diagram. The 
numerator  is computed from the product of vertex factors after expanding  the internal momenta 
into sums of external momenta. When the amplitude is expressed in Pfaffian diagrams, terms with
similar products of $\e_i\.\e_j,  \e_i\. k_j,  k_i\.k_j$ in the numerator are grouped together,
with a coefficient given by  sums
of products of propagators. This reprocessing generally reduces 
the number of terms in a Pfaffian amplitude compared to a Feynman amplitude, though
of course the two amplitudes must be the same after summation. For example, for $n=4$, the number of terms
in the Pfaffian amplitude is 19, far less than the 35 terms appearing in the (color-stripped) Feynman amplitude.

The other advantage of Pfaffian diagrams is that  gauge invariance is simpler and more explicit. 
A single Feynman diagram is never gauge invariant; a sum of several diagrams is needed for the gauge invariance
 of a single external line $i$. To make
it gauge invariant for two $i$'s, a sum of more diagrams is required. The larger the set  of external lines, the more
diagrams must be summed to gain gauge invariance for members of the set. Finally, for the set of all external lines, every Feynman diagram
must be included to attain gauge invariance. For large $n$, this is an astronomical number.

In contrast, there are many small subsets of Pfaffian diagrams which are gauge invariant for every $i$, except two special ones
$\l, \n$ picked out from the very start in constructing the Pfaffian diagrams. 
This property can be used in practice to check and to simplify calculations.
To be gauge invariant for $\l$ or $\n$, again
a sum of all diagrams is required.

Reduced Pfaffians can be expanded into permutation cycles \cite{LY3}, but that causes problems in integration.
In the case of  a bi-adjoint $\f^3$ amplitude, its integrand consists only of simple poles in  
the integration variables $\s_i$, so  residue calculus
can be used to compute the integral easily. The terms in the expanded Pfaffian however contain double  poles, making it much harder to apply residue calculus. It was 
realized in \cite{LY3} in low-$n$ examples that scattering equations can be used to convert double poles into simple
poles, after which integrations can be easily carried out. In this note we find a systematic way to apply this idea  to all $n$
to get rid of the double poles, thereby allowing one to formulate a 
set of rules to compute gluon amplitudes in terms of the Pfaffian diagrams. 

There is some overlap with previous work \cite{pw1,pw2,pw3,pw4,pw5}, but the present approach is more economical and results in less number of terms.
This is so partly because we need not expand our expressions into KK, BCJ, or  Cayley basis, by adding and subtracting additional terms. Integrations are carried out directly in this article.  Further simplification 
is achieved by  a proper  choice
of the three special \M lines $r, s, t$, as well as the two special rows and columns $\lambda, \nu$ in the reduced Pfaffian.
 Note that only  lines $r, s, t, \lambda, \nu$ are chosen, but the specific values of  $\s_r, \s_s, \s_t$
 are left arbitrary to preserve \M invariance of the amplitude.

The expansions of Pfaffians and reduced Pfaffians are reviewed in Section II, both analytically and graphically. 
The method to convert bi-adjoint $\f^3$ integrations into Feynman diagrams is reviewed in Section III, then generalized
to enable it to integrate gluon amplitudes after  double poles are converted into simple poles.
How double poles are converted is discussed in Section IV, analytically and graphically. These prerequisites allow
Pfaffian diagrams and Pfaffian rules to be formulated in Section V. That Section also contains a comparison 
between Pfaffian diagrams and Pfaffian rules with Feynman diagrams
and Feynman rules. Four point amplitudes are computed in Section VI using both Pfaffian diagrams and Feynman diagrams,
to illustrate how Pfaffian rules are used, and how the two ways to compute differ. Finally, Section VII
contains a summary, and Appendix A consists of detailed explanations of the material in Section III.

\section{Expansion of Pfaffians and reduced Pfaffians}
In this section we review the CHY gluon amplitude \cite{CHY1,CHY2,CHY3,CHY4,CHY5}, and the expansions of Pfaffians and reduced Pfaffians \cite{LY3}.

\subsection{The CHY formula}
A color-stripped $n$-gluon scattering amplitude is given by the CHY formula  to be
\be
M^{\a}&=&\(-{1\over 2\pi i}\)^{n-3}\oint_\Gamma\s_{(rst)}^2\(\prod_{i=1,i\not=r,s,t}^n{d\s_i\over f_i}\){I\over\s_{(12\cdots n)}},\labels{ms}\\  I&=&\pf'(\Q),\labels{mg}\ee
where $\s_{(rst)}=\s_{rs}\s_{st}\s_{tr},\
\s_{(12\cdots n)}=\prod_{i=1}^n\s_{i,i\+1}$ with $\s_{n\+1}\equiv\s_1$, and $\s_{ij}=\s_i\-\s_j$. Please note the difference
between $\s_{(12\cdots n)}$ and $\s_{12\cdots n}=\s_{(12\cdots n)}/\s_{n1}$. The product of the former loops back
and the latter does not. Thus $\s_{ij}$ is not the same as $\s_{(ij)}=\s_{ij}\s_{ji}$.

The scattering functions $f_i$ are defined by
\be f_i&=&\sum_{j=1,j\not=i}^n{2k_i\.k_j\over \s_{ij}},\quad (1\le i\le n),\labels{f}\ee
with $k_i$ being the outgoing momentum of the $i$th gluon.
The three special lines $r,s,t$ for the \M constants $\s_r,\s_s,\s_t$ will be referred to as  constant lines,
the rest variable lines.
The reduced Pfaffian $\pf'(\Q)$
is related to the Pfaffian of a matrix $\Q^{\l\n}_{\l\n}$ by
\be \pf'(\Q)={(-1)^{\l+\nu \+n\+1}\over\s_{\l\n}}\pf\(\Q^{\l\n}_{\l\n}\),\quad (\l<\n),\labels{ln}\ee
where $\Q^{\l\n}_{\l\n}$ is obtained from
 the matrix $\Q$ with its $\l$th and $\n$th columns and rows removed. 
The antisymmetric matrix $\Q$ is made up of three $n\x n$ matrices $A, B, C$,
\be
\Q=\bm A&-C^T\\ C&B\\ \em.\labels{QABCD}\ee
The non-diagonal elements of these three sub-matrices are
\be A_{ij}&=&{k_i\.k_j\over\s_{ij}}:={a_{ij}\over\s_{ij}},\quad B_{ij}={\e_i\.\e_j\over\s_{ij}}:={b_{ij}\over\s_{ij}},\nn\\
 C_{ij}&=&{\e_i\.k_j\over\s_{ij}}:={c_{ij}\over\s_{ij}},\quad 
-C^T_{ij}={k_i\.\e_j\over\s_{ij}}={c_{ji}\over\s_{ij}},\quad 
(1\le i\not=j\le n),\labels{ABC}\ee
where $\e_i$ is the polarization of the $i$th gluon, satisfying $\e_i\.k_i=0$. 
The diagonal elements of $A$ and $B$ are zero, and that of $C$ is defined by
\be C_{ii}=-\sum_{j=1}^nC_{ij},\labels{CC}\ee
so that the column and row sums of $C$ is zero. A similar property is true for $A$ if the scattering equations $f_i=0$
are obeyed. This is the case because the integration contour $\Gamma$ encloses these zeros anticlockwise. 

For massless  particles satisfying momentum conservation, the amplitude $M^\a$ is \M invariant, and is 
independent of the choice of
$r, s, t, \l, \n$, as well as the values of $\s_r, \s_s$, and $\s_t$. It is also gauge invariant, in the sense that when any $\e_i$ is replaced by $k_i$, then the amplitude is zero.

\subsection{Expansion of the Pfaffian}
$\pf(\Q)$ can be expanded into a sum over all permutations $\pi\in S_n$ of $n$ objects  \cite{LY3} ,
\be \pf(\Q)=\sum_{\pi\in S_n}(-1)^\pi\Q_I\Q_J\cdots\Q_L,\labels{PF}\ee
where $I, J, \cdots, L$ are cycles of $\pi$,
\be \pi&=&IJ\cdots L=(i_1i_2\cdots i_u)(j_1j_2\cdots j_v)\cdots(l_1l_1\cdots l_w),\nn\\
n&=& u\+v\+\cdots\+w.\labels{CYC}\ee
The cycle factor $\Q_I$ is 
\be \Q_I&=&{\h\Tr(U_{i_1}U_{i_2}\cdots U_{i_u})\over\s_{(i_1i_2\cdots i_u)}}:={\h\Tr(U_I)\over\s_I},\quad   U_a=k_a\e_a-\e_ak_a,\nn\\
 \s_{(i_1i_2\cdots i_u)}&=&\s_{i_1i_2}\s_{i_2i_3}\cdots\s_{i_u\-1,i_u}\s_{i_ui_1},\labels{TR}\ee
when the cycle length $u>1$. Otherwise, if $I=(i)$, then $\Q_I=C_{ii}$. The other cycle factors $\Q_J,\cdots,\Q_L$ are similarly defined.

The signature of a cycle of length $u$ is $(\- 1)^{u\+1}$, hence the signature factor $(\- 1)^\pi$ 
is equal to $(\- 1)^{u\+v\+\cdots\+w+\g}=(\- 1)^{n+\g}$, where $\g$ is the number of cycles in \eq{PF}.

It is important to note that each cycle factor $\Q_I,\cdots,\Q_L$ is gauge invariant in all its external lines.
If $\e_a$ is replaced by $k_a$, every cycle factor vanishes because  $U_a$ is then zero.

When expanded, the trace gives rise to $2^{u}$ terms.  Each is a
product of $u$ factors of $d_{ii'}$, where 
$d_{ii'}$ is either $a_{ii'}, b_{ii'}, c_{ii'}$, or $c_{i'i}$, with  $i,i'$ being consecutive indices in $I$. 
For example, if $I=(246)$, it follows from \eq{ABC} that
\be \Tr(U_2U_4U_6)&=&c_{24}c_{46}c_{62}\-b_{24}a_{46}c_{62}\-a_{24}c_{46}b_{62}\-c_{24}b_{46}a_{62}\nn\\
&+&c^t_{24}a_{46}b_{62}\+a_{24}b_{46}c^t_{62}\+b_{24}c^t_{46}a_{62}\-c^t_{24}c^t_{46}c^t_{62}\nn\\
&:=&(ccc\-bac\-acb\-cba\+c^tab\+abc^t\+bc^ta\-c^tc^tc^t)_{(246)}.\labels{TR3}\ee

There are several general features worth noting about this expansion. 
First of all, the structure of $U_a$ in \eq{TR} dictates
that there is one $\e_a$ and one $k_a$ at each juncture $a$. Thus a $x\.k_a$ for any $x$ must be
following by a $\e_a\.y$ for some $y$, never a $k_a\.y$. In other words, the four letters $a, b, c, c^t$ must be assembled
according to the orders given in Fig.~1, where the circles around $c$ and $c^t$ indicate that $c$ can be followed by
another $c$, and $c^t$ can be followed by another $c^t$.
\bc\igw{2.5}{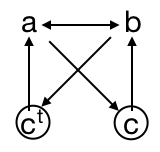}\\ Fig.~1.\quad Allowed orders of $a,b,c,c^t$. For example, $b$ may follow $c$
but never followed by $c$, though it can follow or be followed by $a$. $c$ can be followed by $b$ or another $c$, but never $a$.
\ec

Secondly, cyclical invariance of the trace is preserved and is reflected in terms 2,3,4 of \eq{TR3}, as well as
in terms 5,6,7. Thirdly, $\Tr(U_2U_4U_6)=Tr(U_6^tU_4^tU_2^t)=(-1)^3\Tr(U_6U_4U_2)$ shows that to each term
present there must be another term read backward, with a sign given by $(-1)^u$. This is so for 
the pair of terms
(1,8), (2,5), (3,6), and (4,7). Note also that the number of $b$'s and $a$'s must be the same to keep the total number
of $\e$'s and $k$'s equal.

With these observations, it is easy to write down the $2^u$ terms in any $\Q_I$. A term with the product of $u$ $c$'s
is always present, with a $+$ sign. Starting  from this term, one can obtain other terms by making the replacement
$cc\to -ba$. In addition, we may have to add in terms to make it cyclically invariant, and terms read backward with
a $(-1)^u$ sign. 

The terms constructed this way never contain $c$ and $c^t$ at the same time, but such terms are allowed by Fig.~1
so must be added in. Such terms are of the form $ac^mb(c^t)^n$, or $b(c^t)^nac^m$, for any $m$ and $n$. 
They involve at least four factors, hence absent in \eq{TR4}, but they are generally present.

Using these rules, the 16 terms of a 4-cycle $\Tr(U_I)$ can easily be written down to be
\be &&(cccc)\-(bacc\+accb\+ccba\+cbac)\+(baba\+abab)\+(acbc^t\+cbc^ta\+bc^tac\+c^tacb)\nn\\
&\+&(c^tc^tc^tc^t)\-(c^tc^tab\+bc^tc^ta\+abc^tc^t\+c^tabc^t),\labels{TR4}\ee
where cyclic partners are grouped together in round parentheses, and the second line contains additional terms 
from the first line read backward.

The cycle factor $\Q_I=\Tr(U_I)/\s_I$ is obtained by dividing every
$d_{ii'}$ in $\Tr(U_I)$ by $\s_{ii'}$. According to \eq{ABC}, this simply turns the lower-case symbols into capital symbols:
\be
a_{ii'}\to{a_{ii'}\over\s_{ii'}}&=&A_{ii'},\quad b_{ii'}\to{b_{ii'}\over\s_{ii'}}=B_{ii'},\nn\\
c_{ii'}\to{c_{ii'}\over\s_{ii'}}&=&C_{ii'},\quad c^t_{ii'}\to{c^t_{ii'}\over\s_{ii'}}=-C^t_{ii'}.\labels{ABC2}\ee
Thus from \eq{TR3} we get
\be \Q_{(246)}&=&\[CCC\-(BAC\+ACB\+CBA)\-(C^tAB\+ABC^t\+BC^tA)\+C^tC^tC^t\]_{(246)},\labels{Q3}\ee
and for any 4-cycle $I$ we have
\be \Q_I&=&CCCC\-(BACC\+ACCB\+CCBA\+CBAC)\+(BABA\+ABAB)\nn\\
&+&(C^tC^tC^tC^t)\-(C^tC^tAB\+BC^tC^tA\+ABC^tCc^t\+C^tABC^t)\nn\\
&-&(ACBC^t\+CBC^tA\+BC^tAC\+C^tACB).\labels{Q4}\ee
Note that because of the minus sign in the last expression of \eq{ABC2}, now all terms read backward have the same sign as terms
read forward in both cases.

It would be useful to devise a diagrammatic representation for  the cycle factors $\Q_I$. 
We shall use a heavy dot ($\bullet$) at node $i$ to represent the presence of $\e_i$, and an arrow ($\rightarrow$)
to represent the presence of $k_i$. A line linking two neighboring nodes $i$ and $i'$ contains the scalar product between these two factors divided by $\s_{ii'}$. In this way we arrive at the graphic representation for $A_{ii'}, B_{ii'}, C_{ii'}, C^t_{ii'}$ shown in
Fig.~2(a). Furthermore, a vertical bar inserted at node $i$ represents the factor $U_i$. It contains two terms,
shown in Fig.~2(b). With this notation, \eq{Q3} can be displayed graphically as shown in Fig.~3.

\bc\igw{8}{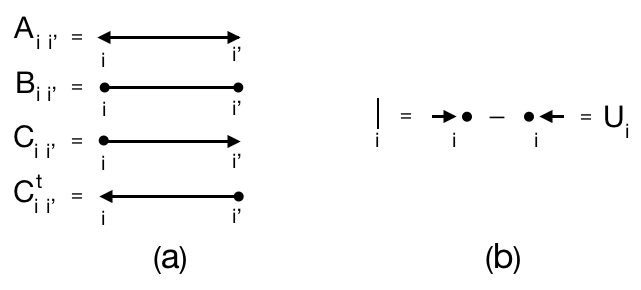}\\ Fig.~2.\quad Graphical representations of $A, B, C, D$ and $U$.\ec

\bc\igw{12}{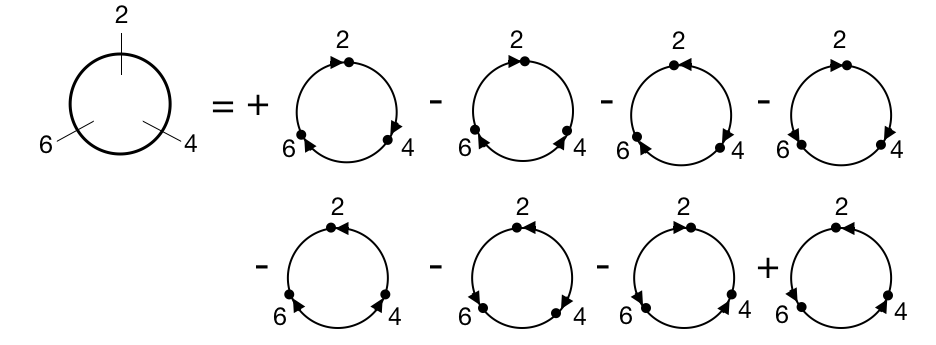}\\ Fig.~3.\quad Graphical representation of $\Q_{(246)}$.\ec

\subsection{Expansion of the reduced Pfaffian}
What is needed in the CHY formula is the reduced Pfaffian defined in \eq{ln}, which can be obtained from
the Pfaffian by the formula 
\be \pf'(\Q)=(-1)^{n\+1}{\p\ \pf(\Q)\over\p a_{\l\n}}.\labels{diff}\ee
Since $a_{\l\n}$ is present in a cycle only when that cycle contains $\l$ and $\n$ in an adjacent  position, differentiation
with respect to $a_{\l\n}$ simply removes that factor and opens up the cycle into a line bounded on the left by
$\e_\l$ and on the right by $\e_\n$.

In this way one gets 
\be
\pf'(\Q)={\sum}_{\pi\in S_{n\-2}}(\- 1)^{\g}W_I\Q_J\cdots \Q_L,\labels{pfp}\ee
where the sum is now taken over all permutations of $n\-2$ numbers, consisting of 1 to $n$ except $\l$ and $\n$. The open cycle factor is
\be W_I={\e_\l U_{i_2}\cdots U_{i_{u\-1}}\e_\n \over\s_{(\l i_2\cdots i_{u\-1}\n)}}.\labels{W}\ee

Since $U_a$ is gauge invariant, every factor of every term in \eq{pfp} is gauge invariant for all external lines,
except lines $\l$ and $\n$. The gauge invariance of these two lines can be seen only after all the terms in \eq{pfp}
are added up.

Expansion for $n=6$ is illustrated graphically in Fig.~4 and Fig.~5. Consider first Fig.~4, ignoring for now the labels 1 and 3, and the
two (out of 13) diagrams within the rectangular box. This is the expansion of $\pf(\Q)$, after the numbers 1,2,3,4,5,6
are distributed in each diagram in all distinct ways, minding cyclic invariance of each cycle. The notation $\Q(m)$ is used
to denote a matrix with $m$ particles, so the original $\Q$ is $\Q(n)$.

Let $(\l, \n)=(1,3)$. To obtain $\pf'(\Q)$ from \eq{diff}, we need to consider those diagrams in $\pf(\Q)$ where the numbers
1 and 3 are adjacent, and differentiate with respect to $a_{13}$ by deleting the link between 1 and 3, then opening up the circle
into a line. The result is shown in Fig.~5. If the pair (1,3) lie in different cycles of a given diagram, all of them must be included,
which is why there are the two `additional diagrams' shown inside the rectangle of Fig.~4.

\bc\igw{11.5}{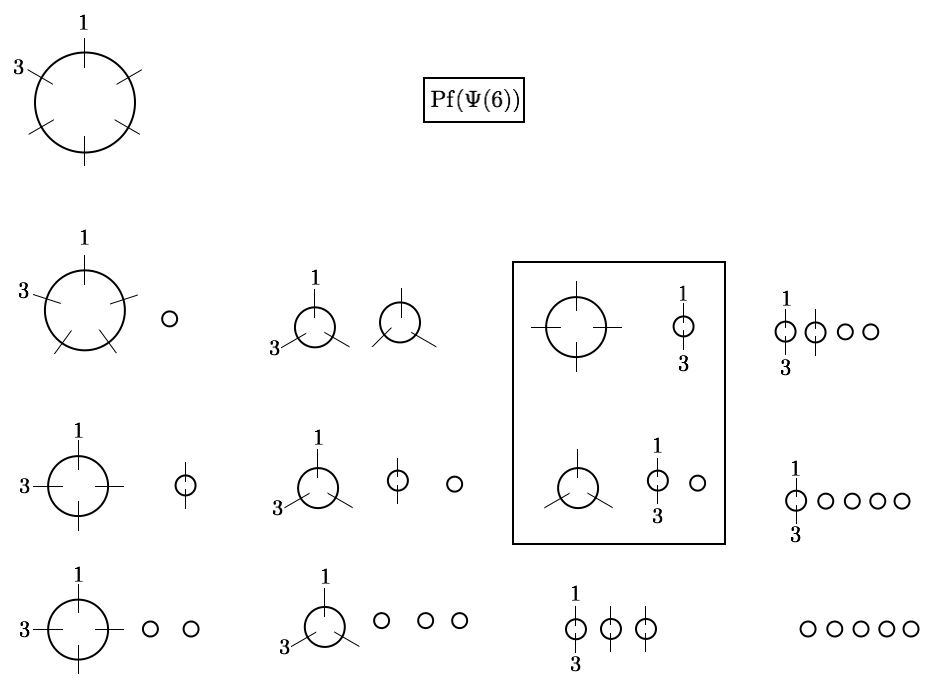}\\ Fig.~4.\quad Graphical expansion for $\pf(\Q)$ for 6 particles.\ec
\vs
\bc\igw{11.5}{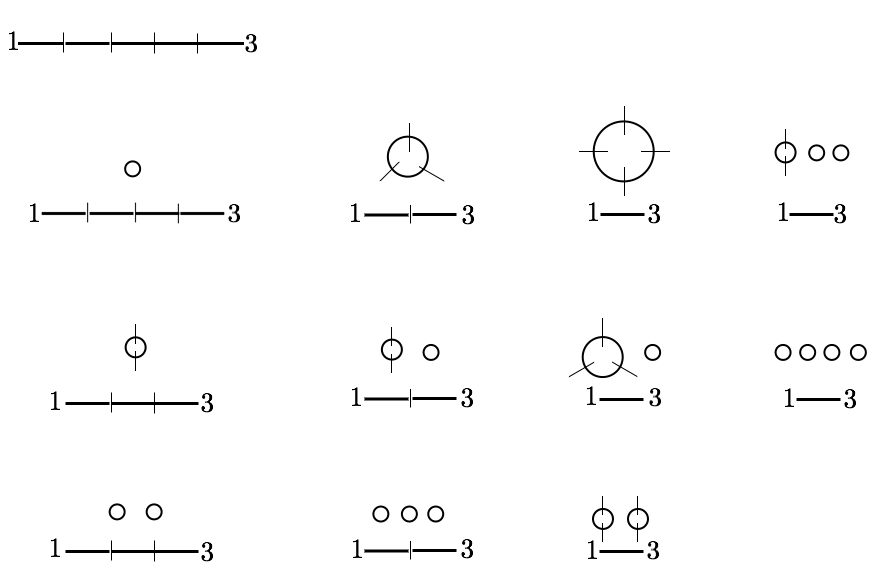}\\ Fig.~5.\quad Graphical expansion for $\pf'(Q)$ for 6 particles when $(\l,\n)=(1,3)$.\ec

\section{Integration and the double pole problem}
Gluon amplitudes can be computed by substituting the expansion \eq{pfp} into \eq{mg}, and carrying out the
integrations over $\s_i$ in \eq{ms}.  Unfortunately,  the presence of double poles makes it difficult to do the integrations directly,
so a better way has to be found.

The bi-adjoint $\f^3$ amplitude is also given by  \eq{ms} but with
\be I={1\over \s_{(\b)}}=\prod_{i=1}^n{1\over \s_{\b_i\b_{i\+1}}},\quad (\b_{n\+1}=\b_1) \labels{mba}\ee
 for some $\b\in S_n$ \cite{CHY2}. Unlike the gluon
amplitude, it has only simple poles
in $\s_i$, so  integrations can be carried out easily using residue calculus, resulting
in a sum of Feynman diagrams determined by the choice of $\b$. Details of how to do that will be reviewed later.

For gluon amplitudes, double poles are present in $\pf'(\Q)$. For example,
$\Q_{(ab)}=\Tr(U_aU_b)/2\s_{ab}\s_{ba}$ contains a double pole in $\s_a$. 
Similarly,   the product of $\s_{ij}$ in any cycle loops back so it also causes a double pole to occur.
Certainly residue calculus can still be used to carry out
integrations in the presence of double poles,
but then derivatives of the rest of the integrand must be computed, making it
very difficult to obtain general rules. This problem is solved in the next section by using
the scattering equations to convert double poles into simple poles, thus allowing a systematic computation whose results can
be formulated into Pfaffian diagrams and Pfaffian rules. As a preparation,
we shall spend the rest of this Section to review how integrations involving only simple poles can be carried out to yield
 Feynman diagrams.
 
 There are many ways to do integrations \cite{CHY3,FG,BBBD2,DG1,BBBD} but we shall follow the method discussed in \cite{LY2}. 
This method differs from the others in that  an explicit choice of the three constant lines $r, s, t$ in \eq{mg} is required,
though the values of their respective \M constants $\s_r, \s_s, \s_t$ remain unspecified.
The latter allows \M invariance of the amplitude to be explicitly verified, by checking the final expression of
 \eq{ms} to be independent
of $\s_r, \s_s$, and $\s_t$.
More importantly, in the case of gluon amplitude, doing the integral with specific choices of $r,s,t$ avoids the
necessity of expanding the integrand in some universal basis by adding and subtracting terms, thereby reducing the
final number of terms in the amplitude. Also, as we shall see,
a judicious
choice of $r, s, t$ can further reduce the total number of terms.

\subsection{Feynman diagrams and their symbolic representations} 
In order to specify which Feynman diagrams emerge from the integrations,
it is convenient to have a symbolic way to describe the diagrams without resorting to pictures. 
The diagrams in question are planar, massless with cubic vertices, and have their external momenta $k_i$
arranged in cyclic order $(123\cdots n)$.
For such diagrams, an internal momentum is always equal to a sum of
some consecutive external momenta $k_i\+k_{i\+1}\+\cdots\+ k_{i\+m}$, thus allowing an internal line or a propagator
to be denoted by $(i,i\+1,\cdots,i\+m)$. Taking $m=0$, the parenthesis symbol can also be used
to denote an external line $i=(i)$. 

This description can be refined to review the structure of the Feynman diagram. If we denote a line $a$ obtained by
 joining (external or internal)
line $b$ and line $c$ as $(bc)$ (Fig.~6), then its momentum is simply the momentum of those in
$b$ and $c$. If $b=(i,i\+1,\cdots,j)$ and $c=(j\+1,\cdots,i\+m)$, then the structure $a=(bc)$
can be revealed by putting a pair of inner parentheses at the appropriate place in
$a=(i,i\+1,\cdots,i\+m)$, to split it up  into $((i,i\+1,\cdots,j)(j\+1,\cdots,i\+m))=(bc)$.

The symbol $(bc)$ can also be understood
to be a vertex where line $b$ merges with line $c$ to form line $a=(bc)$, though this notation is not symmetrical
in the three lines forming the cubic vertex. A symmetrical notation to denote the same vertex is $(a)(b)(c)$.

\bc\igw{2.5}{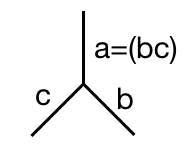}\\ Fig.~6.\quad A vertex at which line $a$ is split up into lines $b$ and $c$. \ec 

A whole Feynman diagram can  be described using these parenthesis symbols. Start from any vertex $v=(R) (S) (T)$ (Fig.~7),
then proceed to split the internal lines repeatedly to expose the inner structure until they cannot be split up anymore. In this
way the whole structure of the diagram is revealed and this can be used as a symbolic representation of the whole diagram.
\bc \igw{6}{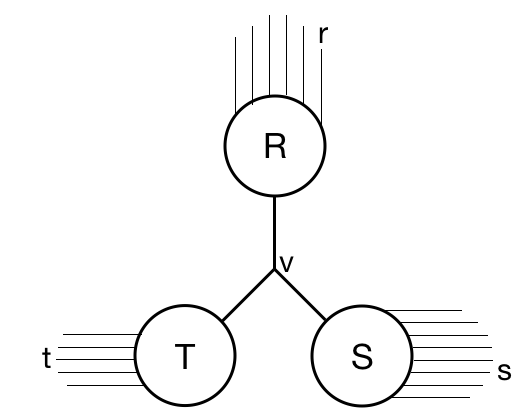}\\ Fig.~7.\quad A Feynman diagram seen at vertex $v$. \ec 
For example, taking $v=u$, Fig.~8(a) can be represented as
\be u=(D)(2)(E)=(BC)(2)(3F)=((A8)(91))(2)(3(45))=(((67)8)(91))(2)(3(45)).\ee
\bc\igw{12}{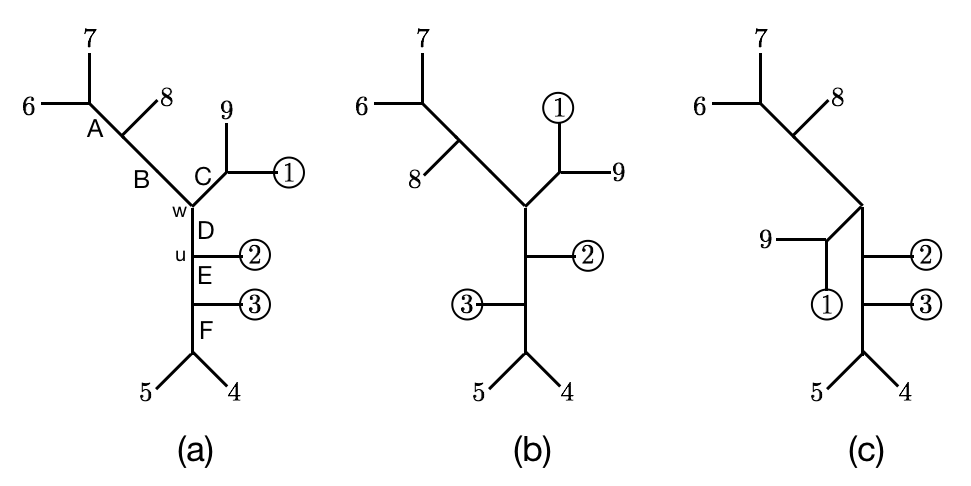}\\ Fig.~8.\quad Three equivalent 9-point Feynman diagrams.\ec 

A Feynman diagram can be represented in many different ways. First of all, a different vertex can be used to
start the split. If we pick $v=w$ instead, then the representation of Fig.~8(a) becomes
\be w=(B)(C)(D)=(A8)(91)(2E)=((67)8)(91)(2(3F))=((67)8)(91)(2(3(45))),\ee
which amounts to a regrouping of the parentheses in the $u$-representation.

The following requirement can be and will be imposed to reduce the number of allowed representations. Take any three external lines $r,s,t$
and require  $r\in R, s\in S$, and $t\in T$. If we apply that to Fig.~8(a) with $(rst)=(123)$, then the $u$-representation is the only one allowed.
This requirement may seem artificial, but when it comes to the CHY formula in \eq{ms}, there are actually
three such special lines, and the integral must depend on them. These three lines must somehow
makes their appearance in the resulting Feynman diagram, and this is how they appear.  

A $\f^3$ Feynman diagram can be drawn in many different ways by flipping lines at vertices. For example,
lines $b$ and $c$ can be interchanged in Fig.~6, and $(bc)$ can be written as $(cb)$.
As a consequence, Figs.~8(a), 8(b), 8(c) are essentially
the same. Flipping corresponds to swapping orders within  parentheses, so  while 
(((67)8)(91))\ (2)\ (3(45)) gives rise to Fig.~8(a),
((8(67))(19))\ (2)\ ((45)3) would give rise to Fig.~8(b), and the two are essentially the same.

Such swapping changes the order of number within parentheses, but one thing never changes: the momentum sum of
all the external lines within each parenthesis remain the same consecutive sum, though the order of the terms
$k_i\+k_{i\+1}\+\cdots\+ k_{i\+m}$ may be interchanged. 
We shall refer to an ordered set of numbers as a {\it consecutive set} if it can be obtained by 
permuting a set of consecutive numbers. For example, (86719) is a consecutive set because if can
be obtained from (67891) by a permutation, but (86729) is not. Thus, no matter how the external lines are flipped, the numbers
within each parenthesis always form a consecutive set. When we refer to a parenthesis from now on, we always
assume the numbers within the parenthesis form a consecutive set. 

In conclusion, a Feynman
diagram can always be represented by a set of  parentheses within parentheses,
such that each parenthesis contains exactly two members, representing the vertex in Fig.~6.
This representation will be referred to as a {\it triple binary split}, indicating how one starts at
a vertex $v$ that divides the diagram into three parts, and how the internal lines in every part
keeps on making binary splits at each subsequent vertices. It will also be simply called {\it pairing},
to conjure up the reverse procedure of constructing a diagram by repeatedly merging a pair of lines
at a vertex.

\subsection{Feynman diagrams of a $\f^3$ amplitude}
Details of how to integrate a bi-adjoint $\f^3$ amplitude is discussed in \cite{LY2} and summarized in Appendix A.
The resulting Feynman diagrams for a given $\s_{(\b)}$ can be obtained  by reversing the discussion of the last subsection.

Starting from a list $(\b)$ of numbers, the first task is to divide the numbers into
three consecutive sets so that $r\in R, s\in S$, and $t\in T$, where $r,s,t$ are the constant lines in \eq{ms}. 
This amounts to picking a vertex $v$ to unravel the diagram.

Next, take each of these three sets and divide it into two consecutive subsets, thereby exposing a vertex at which this
line is split into two lines defined by these two consecutive subsets. Continue this way with every subset
 until no more split
is possible. This then creates a representation of a Feynman diagram. As before, we shall refer this procedure as {\it triple binary splitting}, or simply as {\it pairing}.

The point is, the integral \eq{ms} with $I=1/\s_{(\b)}$ is equal to the sum of the Feynman amplitudes for the
Feynman diagrams created this way by triple binary splits. If such splits cannot be completed, which would
happen if at some point a set can no longer be divided into two {\it consecutive} sets, then the integral is zero. 
For example, $(7546)\to(7(546))$ is a consecutive set that can be so split, but (5746)  cannot be split
into two consecutive sets.
If several
inequivalent splits exist, then the integral is equal to the sum of them. This is so because every
such split defines a dominant integration region in which the integral can be computed and expressed as a Feynman
diagram.

For example, if $(rst)=(123)$ and
$(\b)\=(867192453)$,   then ((8(67))(19))\ (2)\ ((45)3) is the only split possible so the integral is just the Feynman
amplitude for Fig.~8(b). There may be a sign involved which will be discussed later. If $(\b)=(678912345)$, then not only it 
can be paired into (((67)8)(91))\ (2)\ (3(45)) of Fig.~8(a), it can also be paired into any other Feynman diagram, {\it e.g.,}
(((5(67))(89))1)\ (2)\  (34), so the integral is equal to the sum of all  Feynman diagrams whose external lines
are arranged cyclically according to $(\b)$.

\subsection{Simple-poles in two dimensional patterns}
In a bi-adjoint $\f^3$ amplitude, the product of $\s_{ij}$ in \eq{mba} is sequential along a 1-dimensional  list $(\b)$.

For gluon scattering, a more erratic pattern of product  
emerges  after double poles are converted into simple poles. Instead of a 1-dimensional pattern like $(\b)$, the result can usually 
be displayed only in a 2-dimensional connected tree, as illustrated
in Fig.~9(a). The bottom horizontal line of such a pattern will be called a {\it root}, 
and all the other lines with an arrow at the end  {\it branches}. Branches could be horizontal, vertical, or oblique. 
The factors $\s_{ij}$ in the pattern are read circularly from left to right on the root,  and read along the direction of  the arrow on a branch, ending at the arrow.

Integrations can be carried out similar to 1-dimensional patterns, and the result can also be obtained 
from triple binary splits. In this case, when a set is divided into two consecutive sets, each of the
daughter sets must again be a connected tree. 

For example, the tree in Fig.~9(a) stands for $I(\s)=\s_{19}\s_{93}\s_{31}\s_{45}\s_{53}\s_{67}\s_{79}\s_{87}\s_{25}$,
with the short heavy bars dividing it into three parts, containing respectively $(rst)=(123)$.

The result of the integral \eq{ms} with this $I(\s)$ is shown in the rest of Fig.~9, both in terms of Feynman diagrams
and the triple binary split representations of them.

\bc\igw{15}{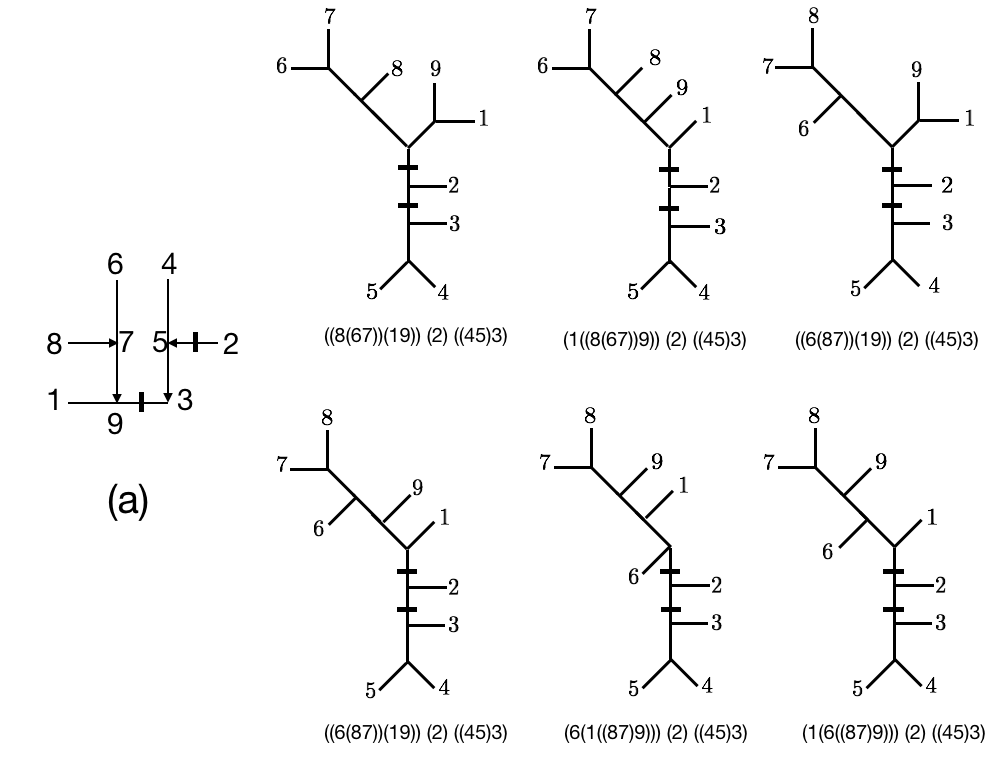}\\ Fig.~9. \quad The 2-dimensional pattern representing an $I(\s)$ is shown  in diagram (a).
Its possible triple-binary splits and the associated Feynman diagrams are shown in the rest of the Figure.\ec

\subsection{Signs and \M constants}
So far we have glossed over two important details: the sign problem, and the  role of the
\M constants $\s_r, \s_s, \s_t$.

Hitherto no distinction is made between $(ij)$ and $(ji)$, because
both represent an internal momentum  $k_i+k_j$. However, $\s_{ij}=-\s_{ji}$, so there should be a sign
difference in \eq{ms} depending on whether $\s_{ij}$ or $\s_{ji}$ appears in $I(\s)$. In 
the case of a 1-dimensional pattern $(\b)$, there should be a minus sign for each parenthesis if the order of the pair  
in the split of $(\b)$ is opposite to the standard order  $(123\cdots n)$.
 For example, in the split $((8(67))(19))\ (2)\ ((45)3)$ of the 1-dimensional pattern $(\b)=(867192453)$,
(8(67)) produces a minus sign because 8 appears before 6 and 7  whereas 8 appears after 6 and 7 in the standard
order. Similarly, ((45)3) also produces a minus sign, and so does the pair (19). Thus the overall sign of this split should be $-$.

The same is true for a 2-dimensional tree, though with a slight complication. Take the upper left Feynman diagram in Fig.~9.
The  split $((8(67))(19))\ (2)\ ((45)3)$ gives rise to an overall $-$ sign just as before. The complication comes in when
this split is written down. It is written as  $((8(67))(19))\ (2)\ ((45)3)$ but not  $(((67)8)(19))\ (2)\ ((45)3)$, for otherwise
a different sign results. The order the numbers appear in the split is determined by the order of the $\s_{ij}$ 
in $I(\s)$, or equivalently in the 2-dimensional tree. In the 2-dimensional tree Fig.~9(a), 8 appears in $\s_{87}$,
so 8 is before 7, which is why it should be (8(67)) and not ((67)8).

Now the question of the \M constants  $\s_r, \s_s,\s_t$. For the amplitude to be \M invariant, they should be cancelled out in the
integral in \eq{ms}, which means that the integration of  $I/\s_{(12\cdots n)}$ must produce a factor proportional to $1/\s_{(rst)}^2$. What makes it interesting and complicated is that this may not happen in every term in the expansion of $\pf'(\Q)$.

As shown in Appendix A, the \M  factors left behind after integration  can be obtained by merging every variable
$\s$'s into the \M constant contained in that part of the triple split.  If we choose $r<s<t$,  the Park-Taylor factor $1/\s_{(12\cdots n)}$ in the
integrand always yields a factor $1/\s_{(rst)}$, making it necessary for $I$ in \eq{ms} to yield another factor
proportional to $1/\s_{(rst)}$. This is clearly so when $I=1/\s_{(\b)}$, but for 2-dimensional patterns, this is not always the case
term by term.

For example, let $(rst)=(123)$ in Fig.~9(a). 
4 and 5 merge into 3 so the factor $\s_{25}$ becomes $\s_{23}$. Thus
the leftover factor of this tree is $1/\s_{(13)}\s_{23}$, and not $1/\s_{(123)}$. 
To render the gluon amplitude \M invariant, another tree(s) with the leftover factor $1/\s_{(13)}\s_{21}$ is needed,
with the {\it same coefficient} and an opposite sign, so that
the combination
\be
{1\over\s_{(13)}}\({1\over\s_{23}}-{1\over\s_{21}}\)={1\over\s_{12}\s_{23}\s_{31}}={1\over\s_{(123)}},\labels{2c}\ee
becomes $1/\s_{(123)}$, making the gluon amplitude  \M invariant. How this can happen and its implication
will be discussed in Sec.~VI.

\subsection{decomposition of 2-dimensional into 1-dimensional patterns}
Integration of a 2-dimensional pattern  can always be converted into a sum of integrations of 1-dimensional 
patterns, by artificially adding and subtracting terms. For example, the upper left corner of Fig.~10 shows a 2-dimensional pattern, whose split into
three groups with $(rst)=(123)$ is shown to its right. This pattern is equivalent to the sum of the three 1-dimensional
patterns below the horizontal line, obtained by inserting 5 in front of every number between 1 and 7. Such 1-dimensional
patterns either generate no allowed split and hence gives a zero integral, or they generate additional splits that cancel one another.

This simple example can be generalized \cite{pw5} to any 2-dimensional tree. The advantage of such a decomposition is that it can be carried out using a set of rules without thinking. 
The disadvantage is that it adds many additional terms that finally cancel one another.
Since we are interested in getting as few terms as possible for the gluon amplitude, we shall not employ such decompositions.

\bc\igw{13}{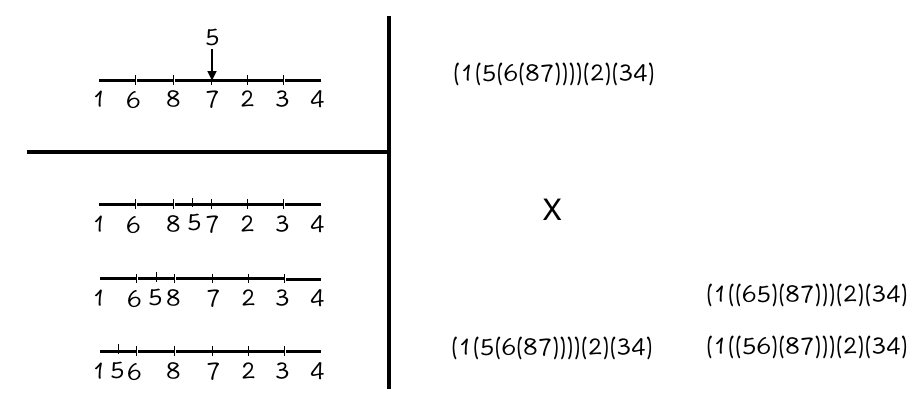}\\Fig.~10.\quad The 2-dimensional pattern above the horizontal line can be
replaced by the sum of the 1-dimensional patterns below. Their respective triple-binary splits are shown
to their right.\ec

\section{Converting double poles of $\pf'(\Q)$ into simple poles}
Double poles are absent in the open cycle factor $W_I$ of \eq{pfp} if $\s_{(\l  i_2\cdots i_{u\-1}\n)}$  contains 
two \M constants to interrupt the looping back of variable $\s$'s. This can be accomplished by assigning  $\l, \n$ to two of the three constant 
 \M lines $r, s, t$, and $\s_\l$ and $\s_\n$  to  two  \M constants.
We shall do so from now, and  shall refer to an open cycle with this assignment as a {\it root}. 

Double poles are also present in the cycle factors $\Q_J,\cdots,\Q_L$ of \eq{pfp}. To get rid of them, the following 
trick can be used to  remove one $\s$ 
factor that causes loop back and double poles to occur.

It is convenient to designate any Pfaffian with $m$ lines as $\pf(\Q(m))$. With this notation, the original Pfaffian is $\pf(\Q(n))$. Let $u$ be the number of nodes in root $I$. The crucial observation is 
that terms in $\pf'(\Q)$ with the same  $u$ 
add up to  a $\pf(\Q(m))$ with $m=n\-u$.
Take Fig.~5 as an example. Counting from left to right, and from
top to bottom, diagram 1 has $u=6$ and diagram 2 has $u=5$. Diagrams 3,4 both have $u=4$ and they sum up to a $\pf(\Q(2))$.
 Similarly,  diagrams 5,6,7 with $u=3$ sum up to a $\pf(\Q(3))$, 
and diagrams 8-12 with $u=2$ sum up to be a $\pf(\Q(4))$. 

Once so grouped, \eq{pfp} can be rewritten as 
\be
\pf'(\Q)=\sum_{m=0}^{n-2} W_I\xi_I\pf(\Q(m)),\ee
provided $\pf(\Q(0))\equiv 1$.  $\xi_I$ is a sign factor that turns out to
be irrelevant and can be ignored.

Concentrate on the first $m$ columns and $m$ rows of  
the $2m\x 2m$ matrix $\Q(m)$, and label them
 as $1,2,\cdots, m$. 
The $a$th column ($1\le a\le m$) contains matrix elements $C_{ba}$ and $A_{ba}$, with  $1\le b\le m$. 
The trick mentioned above consists of adding the remaining $(m\-1)$ columns  to the $a$th column, and the remaining $(m\-1)$ rows  to the $a$th row.
This does not alter $\pf(\Q(m))$, but it changes the $a$th column to
\be C_{ba}\to -\sum_{x\in X}C_{bx},\quad  A_{ba}\to -\sum_{x\in X}A_{bx},\labels{atox}\ee
where $X$ is the set of all numbers {\it not in $\Q(m)$}.
This is a consequence of the scattering equation and the definition of $C_{aa}$ which state that
\be \sum_{b=1}^nC_{ba}=0=\sum_{b=1}^nA_{ba}.\ee
A similar change occurs in the $a$th row. 

This manipulation can be expressed diagrammatically as opening the cycle into a line.
Recall that each node $a$ consists of two terms, with a heavy dot $\bullet$ on one side and an arrow ($\rightarrow$)
on the other side. See Figs.~2 and 3. On the dotted side  $C_{ab}$ or $B_{ab}$ appears, and  on the 
arrowed side it is $A_{ba}$ or $C_{ba}$, with the arrow pointing into node $a$. What \eq{atox} does is to move the
arrow from $a$ to $x$, a point beyond {\it all} the cycles in $\pf(\Q(m))$. See Fig.~11. In this way one breaks the cycle and
removes the double pole.
The opened line begins with   a dot at node $a$ and ends with an arrow at point $x$. It shall be referred to
as a {\it branch}, or the {\it $a$-branch}.

Before we move on there are several details  to be settled first. To start with, there is the question of numerical factors.
A factor $\h$ is present in \eq{TR} for every $\Q_I$, but there are also two terms at every node $a$, corresponding
to reading the scalar products clockwise and anticlockwise. For $x>2$, to every labelling of a cycle there is also another labelling
in the opposite way. 
The end result is that there are no factors of 2, and the branch consists of both orderings, as illustrated in Fig.~11 for $u=3$ and $a=2$.
  
\bc\igw{10}{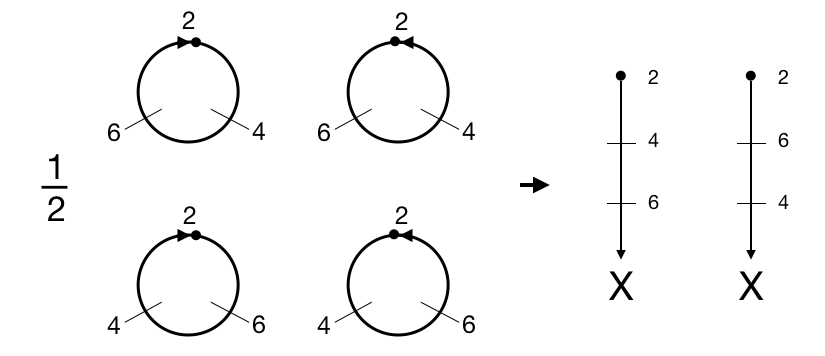}\\ Fig.~11.\quad 3-cycles opening up to branches.\ec

There is a minus sign in \eq{atox} when a cycle is opened up, so there is a total sign of $(-1)^\g$ in a term with $\g$ cycles.
 It cancels a similar sign in \eq{pfp} and turns that equation into
\be
\pf'(\Q)={\sum}_{\pi\in S_{n\-2}}W_I\breve\Q_J\cdots \breve\Q_L,\labels{pfp2}\ee
where $\breve\Q_J$ is the cycle factor $\Q_J$ opened up, namely, with node $a$ at the arrow end replaced by $x\in X$:
\be \breve\Q_J=\sum_{x\in X} {\e_a U_{j_2}\cdots U_{j_v} k_x\over\s_{aj_2}\s_{j_2j_3}\cdots\s_{j_vx}}
:=\breve\Q_{aj_2\cdots j_vx}.\labels{branch}\ee

The replacement in \eq{atox} removes the double pole in the cycle containing $a$, but double poles are still present in
the remaining cycles of $\pf(m)$. Take for example $\pf(\Q(3))$ in diagrams 5,6,7 of Fig.~5. If $a$ is a node in the cycle in diagram 5, opening
that cycle  would remove the double pole in that diagram. However, if $a$ appears in the 1-cycle
in diagrams 6 and 7, then after removing that double pole there are still others left behind.
Diagrams 6 and 7 still contain a $\pf(\Q(2))$ consisting of the 2-cycles of diagram 6 and the two
remaining 1-cycles of diagram 7. One must choose another $a'$ in $\pf(\Q(2))$ to open it up also to remove this double pole.
In other words, \eq{atox} must be used repeatedly to open up all the close cycles.
A term with $\g$ cycles must be opened up $\g$ times, thereby gaining a sign factor $(-1)^\g$ which has already been taken
into account in \eq{pfp2}.

Fig.~12 illustrates what $\pf'(6)$ becomes after the cycles in Fig.~5 are opened up into branches.

\bc\igw{12}{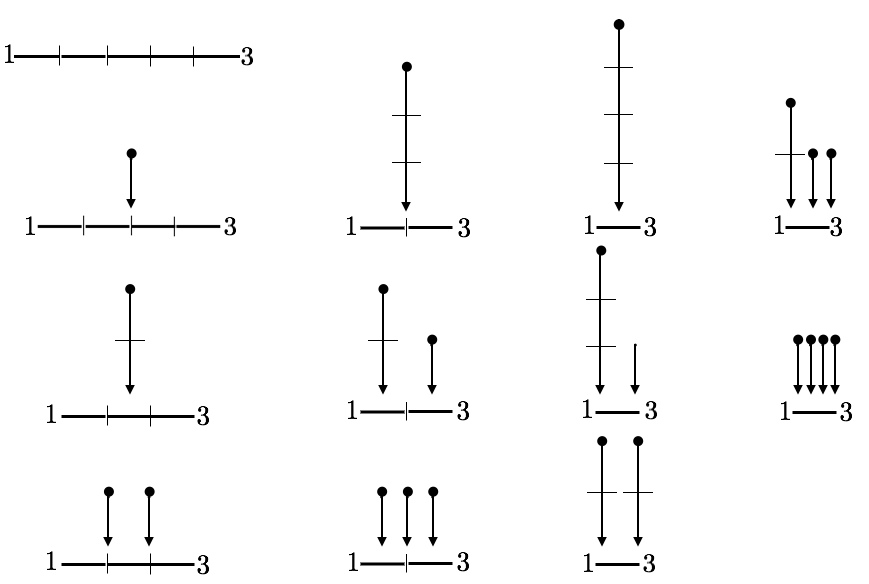}\\ Fig.~12.\quad $\pf'(\Q(6))$ after all cycles are converted to branches. These are 
all the Pfaffian diagrams for $n=6$.\ec

Lastly, there is the question of what the nodes $a, a', \cdots$ are. In principle they are completely arbitrary and their choice should not 
affect the result. To be consistent and to make it easy to remember, we shall choose them in numerical order, namely $a<a'<\cdots$. In practice this means the following.

Choose the smallest number $a$ in $\pf(\Q(m))$ and opens up all the cycles containing that number at that point.
$X$ is the set of nodes in the root so the arrow of this $a$-branch is to be inserted in all possible ways into the root.
Now choose the next smallest number $a'$. If $a'$ is in the $a$-branch, do nothing. If it is in a cycle, then open up that cycle
into a $a'$-branch. Now $X$ consists of everything not in the cycles, so the arrow of the $a'$-branch is to be inserted in all possible
ways into either the root or the $a$-branch. Continue thus until all the cycles are opened up into branches. 
A later (larger) branch  should be inserted into the root and all the earlier (smaller) branches,
but never into a branch to be opened up even later.

So far none of the discussions depend on the choice of the three constant \M lines. We  did decide to put two of them at the two ends
of the root, but we have not yet specified what they are. To simplify the eventual outcome, we shall choose these three 
 constant lines to be
consecutive, and to be 1, 2, 3 without loss of generality. Line 1 is placed at the left end of the root,
and line 3 at the right end. Line 2 can either be on the root or on one of the branches. 

The purpose of this choice is to simplify integration. After insertions, the branches and the root together form a two dimensional tree. For example, Fig.~13 is an $n=9$ tree possessing four branches: a 2-branch, a 4-branch,
a 6-branch, and an 8-branch.
Its two dimensional pattern is essentially that of Fig.~9(a). The only difference is that the paraphernalia of $\pf'(\Q)$
expansion, indicating  $\e$ and $k$ dependences ($\bullet, |$\ ), do not appear in Fig.~9(a), but that does not affect
 how integrations are carried out.
\bc\igw{3.5}{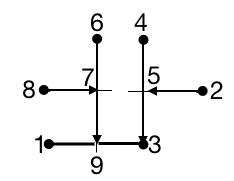}\\ Fig.~13.\quad One possible insertion for an $n=9$ Pfaffian diagram with 4 branches.\ec

Recall from Sec.~IIIC how integrations of 2-dimensional patterns are carried out. The pattern is separated into three parts,
each containing one and only one of the lines 1, 2, 3. Since 2 is sandwiched between 1 and 3, 2 must be the only number
in that group, so effectively we are left with only two connected trees,  one containing
1 and the other containing 3. This greatly simplifies the integration.
Since each part has to be a consecutive set, there must be some demarkation 
point $d$ {\it on the root}, so that one part contains numbers between $3$ and $d$, and the other part contains numbers between $d$
and $n\+1\equiv 1$.
Moreover, 
although 2 is the smallest of all the branches, no branch is allowed to insert 
 {\it on top of} the 2-branch, 
because those numbers above 2 cannot lie in a part containing 1 or 3 without including 2. 
It is still possible to insert
larger branches into the middle of a 2-branch, but never on top of it. 

These remarks can be summarized into a set of \underline{\it branch rules}:
\bn
\i the top (the $\bullet$ end) of a branch always consists of the smallest number in that branch;  
\i a larger branch should be inserted into the root and  smaller branches in all possible ways, but a smaller branch
should never be inserted into a larger branch;
\i no branch is allowed to be inserted on top of the 2-branch.
\en

\section{Pfaffian diagrams and Pfaffian rules}
\subsection{Pfaffian diagram}

The CHY gluon amplitude computed from \eq{ms} and \eq{mg} can be represented by a set of Pfaffian diagrams
with accompanying Pfaffian rules.
 
A Pfaffian diagram is a graphical display of one term $W_I\breve\Q_J\cdots\breve\Q_L$ of the reduced Pfaffian \eq{pfp2}.
Every diagram consists of
one root $W_I$ bounded at two ends  by lines 1 and 3, and any number ($\ge 0$) of branches $\breve\Q_J,\cdots
\breve\Q_L$. 
The total number of nodes in the root and the branches is equal to $n$, the
number of external gluon lines. See Fig.~12 for the $n=6$ Pfaffian diagrams.

Other than 1 and 3, the remaining $n\-2$ numbers are distributed among the root and the
branches in all possible ways, subject to the branch rules at the end of the last section.

Although Pfaffian diagrams are designed to describe the reduced Pfaffian,  they also represent the whole
 gluon amplitude. With the interpretation of
Fig.~2 and with $A_{ii'}, B_{ii'}, C_{ii'}, C^t_{ii'}$ replaced by $a_{ii'}, b_{ii'}, c_{ii'}, -c^t_{ii'}$, the diagram gives directly the numerator of the gluon amplitude.  
The denominator  is a sum of $\f^3$ Feynman diagram propagators  
obtained from the 2-dimensional pattern of the Pfaffian diagram, using the 
triple binary splitting method discussed in Sec.~IIIC.

Remember from the last Section that there is a demarkation point on the root separating 
the tree into a part containing 1 and a  part containing 3. If $2$ appear on the root, then it must be the demarkation point, for otherwise one part would contain two \M constant lines. 
If 2 appears on a branch, 
then situations can occur where roughly `half' the diagrams can be dropped, 
further reducing the final number of diagrams and terms.
This  \underline{\it `half-2 rule'} will be discussed in the next Subsection.

\subsection{Half-2 rule}
This rule  grows out of remarks made in the second half of Sec.~IIIC. Recall  that after  integration,
every variable $\s$ is merged into either $\s_1$ or $\s_3$, depending on which side of the demarkation it belongs to.
What about $\s_2$?

There are three places where $\s_2$ might end up.
If it end up in the middle of the root, then the constant \M factors left over after the integration is $1/\s_{(123)}$,
cancelling the normalization factor in \eq{ms}  to render the amplitude \M invariant. 
If it ends up above 3, as is the case
in Fig.~9 and Fig.~13, then the constant \M factors left over  is $1/\s_{(13)}\s_{23}$. If it ends
up above 1, then the factor is $1/\s_{(13)}\s_{21}$. Since \M invariance of the amplitude 
requires $1/\s_{(123)}$ to be left over, these two cases must combine according to \eq{2c}. 

To be able to combine, the two terms must have equal and opposite coefficients. This requirement provides a tool to check
for errors in computation.

This requirement also allows  all the diagrams proportional to $1/\s_{(13)}\s_{21}$ to be dropped, or
all the diagrams proportional to $1/\s_{(13)}\s_{23}$ to be dropped, because the other half gives only redundant information. In this way,  we can get rid of
half the diagrams, hence the name `half-2 rule'. In practice, the side possessing more diagrams
would be dropped, so the half-2 rule actually ends up with less than half of the diagrams.

\subsection{C-rule}
A special case of the half-2 rule can be implemented easily, leading to a `C-rule' which proves to be very useful in actual calculations.

A branch $\breve\Q_{aj_2\cdots j_vx}$ with $v$ nodes has $2^v$ terms, one of them being 
$C_{aj_2}C_{j_2j_3}\cdots C_{j_vx}\equiv\breve\Q^C_{aj_2\cdots j_vx}$. Diagrammatically it is represented by a branch of arrows. We shall refer
to such a  term as a {\it C branch}, or a $aC$ branch to specify it starts with node $a$.

Consider any diagram $D$ made up of a root and a number of $C$ branches,  but without line 2 present, such as Fig.~14(a). 
Let $D2$ denote the sum of all diagrams with  `possible' and `impossible' 2-insertions into $D$, as in 
Fig.~14(b). Then the C-rule
states that the amplitude of the sum is related to the amplitude of $D$ by the formula
\be M_{D2}=\e_2\.k_{right}\ M_D=-\e_2\.k_{left}\ M_D,\labels{crule}\ee
where $k_{right}$ is the sum of external momenta to the right of the demarkation point (thick vertical line on the root), and
$k_{le\!ft}$ is the sum to the left. This rule, which is a special case of the half-2 rule, will be proven later
and  can be used to simplify actual calculations.
\bc\igw{13}{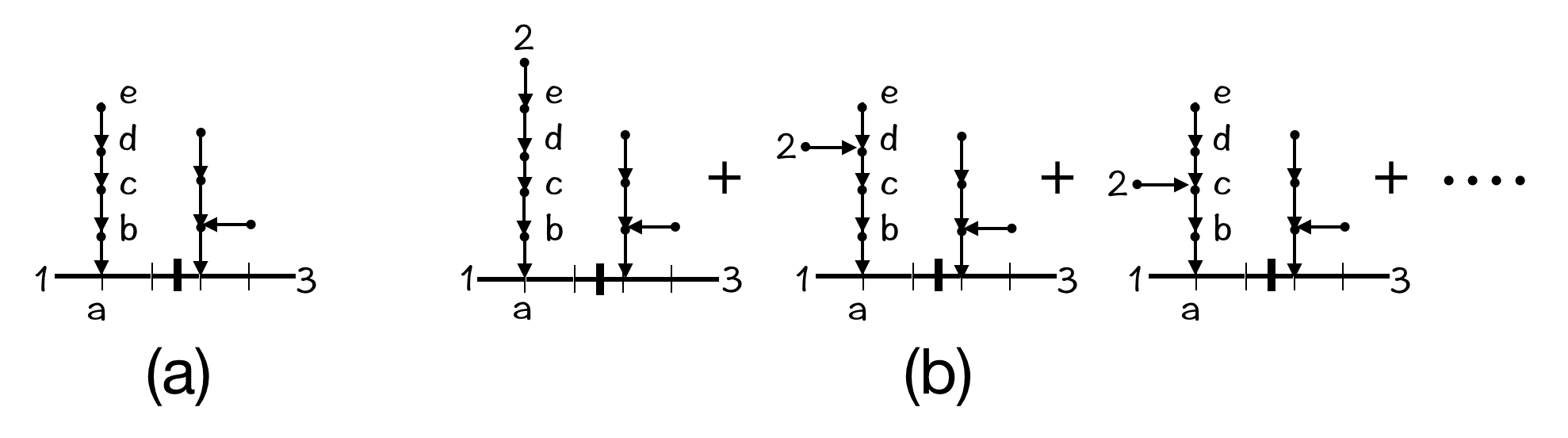}\\ Fig.~14.\quad (a) a diagram with only C branches and no line 2; (b)
`possible' and `impossible' 2-insertions.\ec

`Possible' insertions refer to 2-insertions obeying the branch rule, namely, insertions of the short 2-branch into the root but not
 into other (larger) branches.
`Impossible' insertions refer to  insertions of the short 2-branch into every node of every branch, which are not allowed by the branch rule because 2 is the smallest branch.
Nevertheless, these diagrams are actually allowed if we bend them around as shown in Fig.~15, and interpret
 them as an
insertion of a larger branch into a 2-branch, but not a 2-branch into a larger branch.

\bc\igw{9}{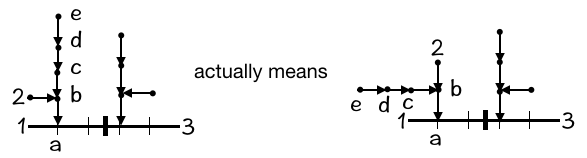}\\ Fig.~15.\quad The `impossible' 2-insertions on the left should be interpreted 
as the allowed insertion on the right.\ec

Proof of the C-rule is simple because $C_{2x}=\e_2\.k_x/\s_{2x}$. After  integration, node $x$ either merges into node 1, or node 3,
depending on which side of the demarkation mark it belongs to. The sum of all the former cases results in an extra
factor $\e_2\.k_{left}/\s_{12}\s_{(13)}$, and the sum of all the latter cases results in an extra $\e_2\.k_{right}/\s_{23}\s_{(13)}$.
\eq{crule} then follows from momentum conservation and \eq{2c}.

\subsection{Comparison between Pfaffian and Feynman diagrams}
The basic components of a Pfaffian diagram are the root and the branches. They can be further decomposed
into four  primitive ingredients: $A, B, C, D$ shown in Fig.~2. The basic ingredients of a Feynman diagram are
the triple gluon vertex, the four gluon vertex, and the propagator, assumed here to be in the Feynman
gauge. On the surface, Pfaffian and Feynman diagrams seem to be completely unrelated.

Pfaffian diagrams are connected together into 2-dimensional patterns according to the branch rules, whereas Feynman
diagrams are assembled
 by connecting up the vertices and the propagators. For a color-stripped Feynman diagram, line
numbers $1, 2, \cdots, n$ are assigned to the external lines in a cyclic order, while in a Pfaffian diagram,
lines 1 and 3 are placed at the two ends of the root, and the other numbers are assigned to the nodes in all possible ways.
Cyclic ordering appears not in the diagrams, but in how the propagators are read out from the 2-dimensional pattern.

A Feynman amplitude is organized by Feynman diagrams. Its denominator is given by the product of 
 internal propagators, and the numerator
is computed from the product of the vertex factors in the diagram. These vertex factors include the polarization vector
$\e_i$ and the momentum $k_i$ of the external lines, but they also contain internal momenta. To get the final
result, internal momenta must be expanded into sums of external momenta. There are always two ways 
connected by momentum conservation to expand every
internal momentum, so the final result can be expressed in many different ways.

In contrast, the Pfaffian amplitude groups together terms with similar numerators, not similar denominators 
as is the case of the Feynman amplitude. Numerators are made up of products of $\e_i\!\cdot\e_j, \e_i\!\cdot k_j,$ and $k_i\.k_j$.
Every $\e_i$ appears once, so there are $n$ $\e$'s in every term, and also $(n\-2)$ $k$'s. The product 
can be read off directly from the Pfaffian diagram with the help of Fig.~2, after dropping the $\s$-factors. 
The denominators are made up of sums
of products of internal propagators, read off from the 2-dimensional pattern using triple binary splits.
Expansion of internal momenta is avoided because Pfaffian amplitudes have no internal momenta. Since the Pfaffian
is symmetrical in all the external momenta, the Pfaffian amplitude gives rise to a numerator which is as symmetrical as possible
in the external data.
 
Gauge invariance demands the gluon amplitude to be zero with the replacement $\e_i\to k_i$, for every $i$. This is never the 
case for any single Feynman diagram. Given an $i$, several diagrams must be summed to ensure its invariance, and the set of diagrams
needed depends on what $i$ is. The amplitude is fully gauge invariant only when all the Feynman diagrams are summed.

For Pfaffian diagrams, gauge invariance for $i=1$ and 3 again needs a summation over all the diagrams. 
However, for the other lines, every subset with a definite root configuration is already gauge invariant. The basic reason
for that is because every $U_i$ is gauge invariant.

If line $i (\not=1,3)$ situates in the root, its gauge invariance follows from the fact that only $U_i$ enters into the root factor $W_I$.
This is also the case if $i$ resides in a cycle of $\pf'(Q)$ before they are opened up into branches. In fact, in that case
every term in \eq{pfp} is gauge invariant.

However, when cycles are opened up into branches, different cycles within the same $\pf(m)$ combine to eliminate the double
pole, so individual branches are no longer gauge invariant. It is only when all the branches associated with the same
$\pf(m)$ added together that becomes gauge invariant. In other words, only the subsets with a given root configuration.

An important attribute of the Pfaffian amplitude is that it usually has less terms than the Feynman amplitude.
This is so  because there is no need for internal momenta expansion,
and because many diagrams can be discarded using the half-2 rule, the C-rule, and the integration rule. See the next Section for a concrete example.

The rest of this Subsection is devoted to a  diagrammatic comparison between Feynman diagrams
and Pfaffian diagrams. Let us start with $n=3$.

Feynman amplitude from the triple gluon vertex has three terms that can be represented in three separate diagrams in
Fig.~16. As before, $\bullet$ represents $\e$, and the box $\Box$ represents the difference of two $k$'s.
\bc\igw{11}{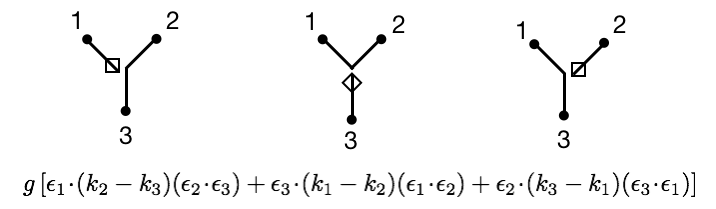}\\ Fig.~16.\quad The three terms of a triple-gluon vertex.\ec

Using momentum conservation and setting $g=\h$, these three terms can be transformed into Fig.~17, where 
as usual an arrow
at node $i$ means $k_i$.
\bc\igw{9}{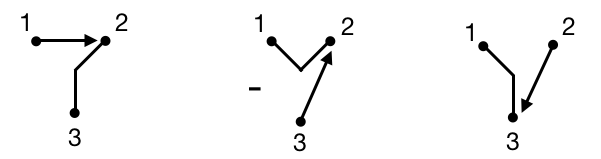}\\ Fig.~17.\quad These three diagrams are equivalent to the three diagrams in Fig.~16.\ec
In this form they can be seen to be identical to the Pfaffian diagrams shown in the second line of Fig.~18.
\bc\igw{9}{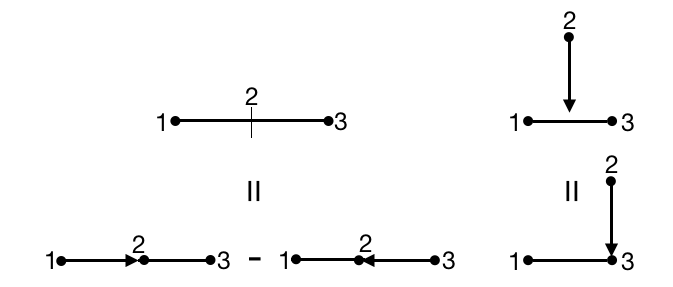}\\ Fig.~18.\quad Pfaffian diagrams for $n=3$.\ec
Thus at least for $n=3$,  Pfaffian diagrams are essentially
 the same as Feynman diagrams after momentum conservation
is used.
For a larger $n$, it is almost impossible to do the expansion of internal momenta and vertex products by hand,
and then to arrange them into Pfaffian diagrams in a systematic way, though one can still see a a connection between the two. For example, take the Pfaffian diagram in Fig.~13, which has the 2-dimensional pattern of Fig.~9(a) and contains the Feynman diagram Fig.~8(b). 
This Feynman diagram contains 7 triple-gluon vertices and therefore $3^7$ terms, even before internal momenta are expanded into sums
of external momenta. One of these terms is shown in Fig.~19(a), where a box $\Box$ represents the difference of two
momenta as before. It can be seen that it does contain the Pfaffian diagram term shown in Fig.~19(b) after its external momenta are suitably replaced by its external momenta by using momentum conservation.

\bc\igw{10}{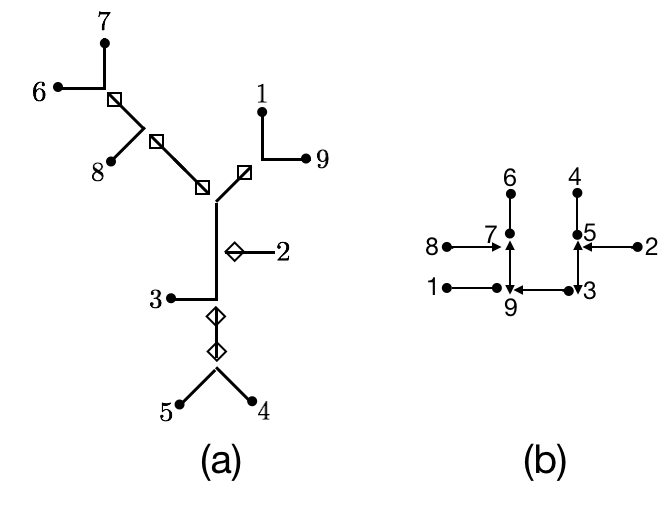}\\ Fig.~19. (a) represents a term of the 9-point Feynman diagram. It contains the
Pfaffian diagram shown in (b). \ec

\section{Four-point amplitude}
Its six Pfaffian diagrams  are shown in Fig.～20, where columns I, II, III contain 
roots with 4, 3, 2 nodes.  Since 1,2,3 are constant lines, $\s_4$ is the only variable to be integrated. 

\bc\igw{14}{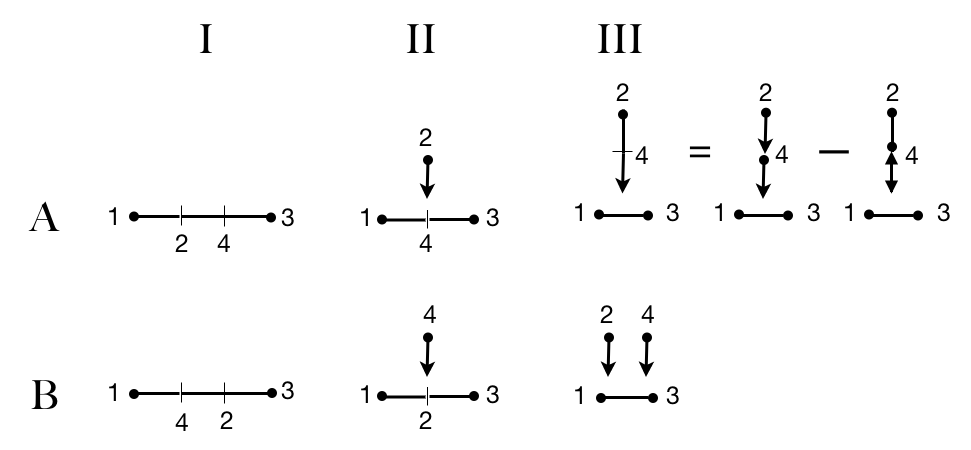}\\ Fig.~20.\quad Four-point Pfaffian diagrams\ec

IA produces the triple binary split  (1)(2)(43)  and IB the split (14)(2)(3), hence
\be {\rm IA}=-{\e_1U_2U_4\e_3\over s_{34}},\quad {\rm IB}=-{\e_1U_2U_4\e_3\over s_{41}},\labels{4I}\ee
where $s_{34}=(k_3\+k_4)^2$ and $s_{41}=(k_4\+k_1)^2$.

Next consider IIA, which consists  of 2-insertions into the root.
The two allowed pairings of the root are
(14)(3), with an demarkation mark between 3 and 4, and (1)(43), with an demarkation mark between 1 and 4.
Using the C-rule, the former gives rise to the amplitude
$-(\e_1U_4\e_3/s_{41})(+\e_2\.k_{right})=-(\e_1U_4\e_3/s_{41})\e_2\.k_3$, and the latter gives rise to  $-(\e_1U_4\e_3/s_{34})(-\e_2\.k_{left})=(\e_1U_4\e_3/s_{34})\e_2\.k_1$. The total contribution to IIA is
therefore
\be {\rm IIA}=\e_1U_4\e_3\({\e_2\.k_1\over s_{34}}-{\e_2\.k_3\over s_{41}}\).\labels{4IIA}\ee

Allowed pairings in IIB depends on where the 4-branch inserts. There is no valid pairing for an insertion at 2, hence no contribution.
Insertion at 1 produces the pairing (41)(2)(3), and insertion at 3 produces the pairing (1)(2)(43). Hence
\be {\rm IIB}=\e_1U_2\e_3\({\e_4\.k_1\over s_{41}}-{\e_4\,k_3\over s_{34}}\).\labels{4IIB}\ee
Note that \eq{4IIA} and \eq{4IIB} differ only by the interchange of 2 and 4, just like the diagrams, although 
completely different methods are used
to arrive at the two results.

Finally we come to the third column of Fig.~20. Expand the 2-branch $\breve\Q_{24x}$ in IIIA into two terms, $C_{24}C_{4x}-B_{24}A_{4x}$, the first constitutes a C branch. The combination of the first term and IIIB   gives rise to all (possible and impossible) 2-insertions of the sub-diagram $D=W_{13}\breve\Q_{4x}$, namely, 
diagram IIIA with the 2-branch removed. The pairing is (41)(3) if the 4-branch is inserted at 1, and (1)(43) if it is inserted at 3.
Using the C-rule on each of these two diagrams, the amplitude becomes 
\be {\rm III}'=\e_1\.\e_3\({\e_4\.k_1\ \e_2\.k_3\over s_{41}}-{\e_4\.k_3\ \e_2\.k_1\over s_{34}}\).\ee
Lastly, there is the  contribution from  $- b_{24}a_{4x}$  in IIIA, where half-2 rule applies, so we have to consider
only the insertion of the 2-branch at 3. This produces a pairing (1)(2)(43), hence a factor $-1/s_{34}$ which cancels the
factor $-a_{43}=-1/2s_{34}$, leaving behind
\be {\rm III}''=\h \e_1\.\e_3\ b_{24}=\h \e_1\.\e_3 \ \e_2\.\e_4,\ee
accounting for part of the four-gluon vertex contributions in Feynman diagrams. The total contribution from column III is therefore
\be {\rm IIIA+IIIB}=\e_1\.\e_3\({\e_4\.k_1\ \e_2\.k_3\over s_{41}}-{\e_4\.k_3\ \e_2\.k_1\over s_{34}}\)+ \h \e_1\.\e_3 \ \e_2\.\e_4.
\labels{4III}\ee

Since each $U_i$ has two terms, there are 8 terms in \eq{4I}, 4 terms each in \eq{4IIA} and \eq{4IIB}, and 3 terms in 
\eq{4III}, making a total of 19 terms.

For lines 2 and 4, gauge invariance shows up in each IA, IB, IIA, IIB,  and IIIA+IIIB, because each of them has a different
root configuration.

In order to compare and to verify, let us compute the Feynman amplitude from Fig.~21.
\bc\igw{8}{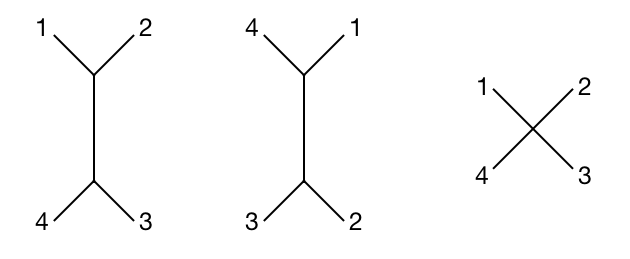}\\ Fig.~21.\quad Four-point Feynman diagrams.\ec
The $s_{12}$-channel amplitude consists of 16 terms,
\be {g^2\over s_{12}}&&\[\e_1\.\e_2\ (k_1-k_2)+\e_1\.(2k_2+k_1)\ \e_2-\e_2\.(2k_1+k_2)\ \e_1\]\cdot\nn\\
&&\[(k_3-k_4)\ \e_3\.\e_4+\e_4\ \e_3\.(2k_4+k_3)-\e_3\ \e_4\.(2k_3+k_4)\],\ee
the $s_{41}$-channel amplitude contains another 16 terms,
\be {g^2\over s_{41}}&&\[\e_2\.\e_3\ (k_2-k_3)+\e_2\.(2k_3+k_2)\ \e_3-\e_3\.(2k_2+k_3)\ \e_2\]\cdot\nn\\
&&\[(k_4-k_1)\ \e_4\.\e_1+\e_1\ \e_4\.(2k_1+k_4)-\e_4\ \e_1\.(2k_4+k_1)\],\ee
and the four-gluon contact amplitude contains another 3 terms,
\be g^2\[2(\e_1\.\e_3) (\e_2\.\e_4)-(\e_1\.\e_2) (\e_3\.\e_4)-(\e_2\.\e_3) (\e_4\.\e_1)\],\ee
making a total of 35 terms, compared to the 19 terms in the Pfaffian amplitude. With $g=\h$, I have explicitly 
verified that the Pfaffian amplitude is identical to the Feynman amplitude.

\section{Conclusion}
The CHY formula for gluon amplitude in any dimension 
can be expressed as a sum of Pfaffian diagrams, whose
ingredients are roots and branches. Each diagram is composed of several branches grown on one root,
from which the amplitude can be read out. Pfaffian diagrams may be considered as a symmetric rearrangement
of the terms in Feynman diagrams, after internal momenta are  expanded into sums of external momenta
and products of vertex factors are computed.
This processing generally renders Pfaffian diagrams to carry less terms  than Feynman diagrams. 
Pfaffian diagrams have the further advantage that gauge invariance is explicit in any subset of diagrams with
a fixed root configuration, for all lines but two. This is a property not share by the Feynman diagrams.

There are three critical technical bottlenecks to get through before these Pfaffian diagrams
can be obtained. Double poles
in the expansion of the reduced Pfaffian must be systematically converted to simple poles. Integration
in the resulting simple poles must be efficiently carried out and formulated into simple rules.
Finally, the three \M constant lines must be suitably chosen to simplify the Pfaffian rules and to minimize
the number of terms present. 

Considerable simplification in the computation occurs in four dimensions \cite{zY}.
The technique to get rid of double poles can also be used on other CHY theories where reduced Pfaffians are present \cite{CHY3,CHY4,CHY5},
though the method of integration must be adjusted because of the modification or the absence of the Parke-Taylor
factor in these other theories. 

I am grateful to Song He for interesting discussions.

\newpage
\appendix
\section{Triple binary split and integration}
This Appendix serves to review  how the triple binary splitting procedure can be used to evaluate the integral
\be
M&=&\(-{1\over 2\pi i}\)^{n-3}\oint_\Gamma\s_{(rst)}^2\(\prod_{i=1,i\not=r,s,t}^n{d\s_i\over f_i}\){I(\s)\over\s_{(12\cdots n)}},\labels{ams}\ee
and used to display the result of the integration. More details can be found in \cite{LY2}.

As in \eq{ms}, the Parke-Taylor factor is defined by
$\s_{(12\cdots n)}=\prod_{i=1}^n\s_{i,i\+1}$, the quantity $I(\s)$ consists of a product of $n\+1$ factors of $\s_{ij}^{\-1}$
 that can be displayed in a connected planar tree,
and the scattering functions are defined by
\be f_i&=&\sum_{j=1,j\not=i}^n{2k_i\.k_j\over \s_{ij}}.\quad (1\le i\le n).\labels{af}\ee
The result of the integral can be obtained from the triple binary splits of $I(\s)$, as explained in Section IIIC and briefly
reviewed below.

A consecutive set is a set of consecutive numbers possibly with a rearranged order.
Thus (86791) is a consecutive set because it is a rearrangement of (67891), but (18675) is not.
A binary split is a partition of  a consecutive set into two consecutive subsets, {\it e.g.,} $(86791)\to((867)(91))$.
Triple split refers to separating the tree for $I(\s)$ into three connected trees of consecutive sets  $R, S, T$, each
containing one and only one of the three constant lines  $r, s, t$.
For example, if $(rst)=(123)$, the triple split of 
$I(\s)=\s_{19}\s_{93}\s_{31}\s_{45}\s_{53}\s_{67}\s_{79}\s_{87}\s_{25}$ of Fig.~9(a)  
results in $R=\(18679\), S=\(2\), T=\(453\)$. The separation is  marked
by two heavy vertical bars in the diagram.  

A complete triple binary split puts $I(\s)$ into a set of parentheses inside parentheses, each representing an
internal line. The integral corresponding to such a split will be shown to be
 $\pm\prod_P(1/k_P^2)$, where $k_P$ is the sum of all external momenta
in parenthesis $P$, with the product  taken over all the parentheses of $I(\s)$. 
The sign is determined by the ordering of indices in $I(\s)$, as explained in Section IIID.
If there are
several ways to do the triple binary splits, then the integral is equal to the sum of all these splits. 
For example, the tree Fig.~9(a) gives rise to splits and  Feynman
diagrams shown in the rest of Fig.~9.

Let us proceed to discuss why a triple binary split can tell us how to 
integrate  \eq{ams}, and why the result can be read out from the split.

There is no way to evaluate the integral directly from the contour $\G$ because
solutions of $f_i(\s)=0$ are unknown for large $n$.
A way out is to distort  $\G$
to surround and to evaluate at the simpler and  explicit singularities of $I(\s)/\s_{(12\dots n)}$.
In order to avoid dealing with
multi-dimensional topology needed for a multi-variable complex integration, 
the integrations will be carried out one at a time, each time  distorting the contour
away from one $f_i=0$.

The singularities of $I(\s)/\s_{(12\cdots n)}$  occur in regions where a number  of $\s_{ij}$ are 
small, and it turns out that these regions are fixed by the consecutive sets of $I(\s)$. 
There are $n\-3$ such parentheses in $I(\s)$,
 each for one of the $n\-3$ integrations in \eq{ams}, each producing one of the $n\-3$
 propagators of the Feynman diagram after the integration.
 
 There are two kinds of parentheses in $I(\s)$, seeded and unseeded. 
 A seeded parenthesis is one that carries a number $c$ (the `seed') whose $\s_c$ is not an integration variable.
 An unseeded parenthesis is one in which all $\s_i$ are integration variables.
 The parentheses
 for $R,S,T$ are seeded with seeds $r,s,t$, but many of their sub-parentheses  are  unseeded. We shall see that 
 a new seed is produced each
 time an integration is carried out, thereby turning some unseeded parentheses into seeded ones.
This makes it possible for the the following method which works only for seeded parentheses to be used repeatedly 
to carry out all the integrations in \eq{ams}.

Consider the integral
\be
K_D=\(-{1\over 2\pi i}\)^d\oint_{\Gamma_D}\(\prod_{i\in D,i\not=c}{d\s_i\over f_i}\){I_D(\s)\over \s_D}\labels{aK}\ee
for a seeded parenthesis $D$ in $I(\s)$. It contains $d\+1$ numbers including the seed  $c$. 
The contour $\G_D$ encircles 
$f_i=0\ \forall i\in D\backslash c$ counter-clockwise.   
$I_D(\s)$ is the portion of $I(\s)$ of the form $\prod_{i,j\in D}(1/\s_{ij})$, and 
$\s_D$ is the portion of the Parke-Taylor factor of the form $\s_D=\prod_{i\in D,i\+1\in D}\s_{i,i\+1}$.
Both contain $d$ factors.

Pick any $p\in D$ which is different from $c$.
Distort the contour $\G_D$ away from $f_p=0$ to surround the singularities of $I_D(\s)/\s_{D}$.
Make the change of variables
$\s_{ij}=\e\s'_{ij}$ for $i,j\in D$, with $\e=\s_{pc}$ so that $\s'_{pc}=1$.
The integration measure is then converted to
\be \prod_{i\in D,i\not=c}d\s_i&=&\e^{d\-1}d\e\prod_{i\in D,i\not=c,p}d\s'_i.\ee
Since the explicit singularities  lie in the small $\s_{ij}$ region, to evaluate the integral we need to consider
what happens to the integrand in the $\e\to 0$ limit.

In that limit, $f_i(\s)\to f^D_i(\s')/\e$, where
\be
f^D_i=f^D_i(\s')=\sum_{j\in D, j\not=i}{2k_i\.k_j\over \s'_{ij}}.\labels{afd}\ee
These functions satisfy the sum rules
\be \sum_{i\in D}f^D_i&=&0,\nn\\
\sum_{i\in D}f^D_i\s'_i&=&\h\sum_{i,j\in D. i\not=j}{2k_i\.k_j\over\s'_{ij}}\(\s'_i\+\s'_j+\s'_{ij}\)=\(\sum_{i\in D}k_i\)^2:=k_D^2.
\labels{asum}\ee
Since  $\G_D$ still surrounds $f_i=0$ for
$i\in D, i\not=c,p$, or equivalently $f^D_i=0$ in the $\e\to 0$ limit,   \eq{asum} implies $f^D_c+f^D_p=0$ and $f^D_c\s'_c+f^D_p\s'_p=k_D^2$ on the contour, hence
\be
f^D_p=-f^D_c={k_D^2\over\s'_{pc}}=k_D^2\labels{afp}\ee
because $\s'_{pc}=1$.

As $\e\to 0$,  the factors $I(\s)\to \e^dI_D(\s')$, and $\s_D\to\e^d\s'_D$, because both  contain $d$ factors of $\s_{ij}$. In toto,
the $\e$ dependence in the integrand of \eq{ams} becomes $d\e/\e$, showing the presence of a simple pole 
which allows the integral to be evaluated by residue calculus. The rest of the integrand is evaluated at $\e=0$,
in particular,  $\s_i=\s_c$ for all $i\in D$.

If the $d\+1$ numbers in $D$ did not form a consecutive set, then $I_D(\s)$ would yield a smaller power of $\e$ than
$\e^d$, causing the simple pole to disappear and the integration to be zero. This is why the dominant integration
region at small $\s_{ij}$'s must come from a consecutive set.

At $\e=0$, $\e f_p$ becomes $f^D_p=k_D^2$.
If the parenthesis $D$ splits into two parentheses $D_1$ and $D_2$,  with $d_1\+1$ and $d_2\+1$
members respectively, then one of them, say $D_2$,  must contain $c$. We will choose $p$ to be situated in $D_1$
so that both $D_1$ and $D_2$ are seeded parentheses, with seeds $p$ and $c$ respectively. 
Moreover, $I_D(\s')=I_{D_1}(\s')\s'_{xy}I_{D_2}(\s')$, and $\s'_{D}=\s'_{D_1}\s'_{ab}\s'_{D_2}$, where $x\in D_1$
is the point next to $y\in D_2$ on the tree, and $a\in D_1$ is the number next to the number $b\in D_2$.
See Fig.~22.
\vs
\bc\igw{4.5}{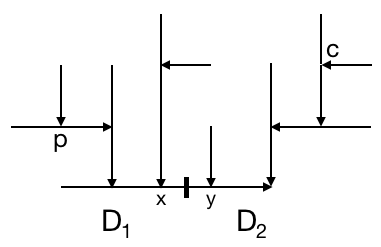}\\ Fig.~22.\quad A seeded parenthesis $D$ with seed $c$ is split into two seeded parentheses
$D_1$ and $D_2$ with seeds $c$ and $d$ respectively.\ec

The remaining $d\-1=d_1\+d_2$ integrations in \eq{aK} can be factorized into
\be
K_D&=&K_{D_1}{1\over k_D^2}K_{D_2},\qquad {\rm with}\nn\\
K_{D_1}&=&\(-{1\over 2\pi i}\)^{d_1}\oint_{\Gamma_{D_1}}\(\prod_{i\in D_1,i\not=c}{d\s_i\over f^D_i}\){I_{D_1}(\s)\over \s_{D_1}},\nn\\
K_{D_2}&=&\(-{1\over 2\pi i}\)^{d_2}\oint_{\Gamma_{D_2}}\(\prod_{i\in D_2,i\not=p}{d\s_i\over f^D_i}\){I_{D_2}(\s)\over \s_{D_2}}.\labels{afactor}
\ee
 If $d_i=0$, then
$K_{D_i}=\pm1$, with the sign determined by the method in Section IIId. If $d_i>0$, then
the treatment for \eq{aK} can be repeated on $K_{D_i}$ to factorize it into two $K$-integrals for its sub-parentheses.
This procedure can be repeated over and over again until all the integrations of \eq{aK} are carried out.
According to \eq{afactor}, each integration produces a propagator $1/k_D^2$ appropriate to that parenthesis,
so the result of all integration would produce a product of $d$ propagators for the $d$ parentheses in $D$, up to a sign.

In getting to \eq{afactor}, the factors $\s'_{xy}$ in $I_D(\s')$ and the factor $\s'_{ab}$ in $\s'_D$ have been set equal
to $\s'_{pc}=1$, in anticipation of subsequent integrations of $K_{D_1}$ and $K_{D_2}$ which would put all the $\s'_i$
for $i\in D_1$ to be equal to $\s'_p$, and all the $\s'_j$ for $j\in D_2$ to be equal to $\s'_c$.

Let us now apply this method to \eq{ams}. As a result of triple splitting, $I(\s)=I_R(\s)I_S(\s)I_T(\s)\r(\s)$ and
 $\s_{(12\cdots n)}=\s_R\s_S\s_T\r'(\s)$. What $\r(\s)$ and $\r'(\s)$ are depend on $r,s,t$ and where they are
 situated in the tree. Consequently,
 \be M=K_RK_SK_T{\s^2_{(rst)}\over \r(\s)\r'(\s)},\labels{afactor3}\ee
so up to a sign and the $\r_T={\s^2_{(rst)}/ \r(\s)\r'(\s)}$ factor, the value of $M$ for any triple binary split
is given by  the product of the $n\-3$ propagators determined by the parentheses of that split. At the
end of all integrations,  $\s_i=\s_r\ \forall i\in R,\ \s_j=\s_s\ \forall j\in S,\ \s_k=\s_t\ \forall k\in T$, so
$\r_T$ can be evaluated with these substitutions.

Let us see what the left-over factor $\r_T$ is for the Pfaffian diagrams, where $(rst)=(123)$, with 1 at the left end
of the root and 3 at the right hand of the root. In that case, $\s_{(rst)}=\s_{(123)}, \r'(\s)=\s_{(123)}$.
The value of $\r(\s)$ 
is equal to $\s_{(123)}$ if 2 is on the root, and is equal to $\s_{(13)}\s_{23}$ if the 2 appears on a branch, and if
the 2-branch is to the right of the demarkation point. It is equal to $\s_{(13)}\s_{21}$ if 
the 2-branch is to the left of the demarkation point.

$R,S,T$ are seeded consecutive sets, so $K_R,K_S,K_T$ in \eq{afactor3} can be evaluated using \eq{aK}
and \eq{afactor}. The result is given  by the triple binary split algorithm, with signs and \M
factors determined in the way explained in Sec.~IIID.

\vs


\begin{thebibliography}{9}
\bibitem{CHY1} F. Cachazo, S. He, and E.Y. Yuan, Phys.~Rev.~D {\bf 90} (2014) 065001 [arXiv: 1306.6575].
\bibitem{CHY2} F. Cachazo, S. He, and E.Y. Yuan,	Phys.~Rev.~Lett. {\bf 113} (2014) 17161 [arXiv: 1307.2199].
\bibitem{CHY3} F. Cachazo, S. He, and E.Y. Yuan,	JHEP {\bf 1407} (2014) 033 [arXiv: 1309.0885].
\bibitem{CHY4} F. Cachazo, S. He, and E.Y. Yuan,	JHEP {\bf 1501} (2015)121 [arXiv: 1409.8256].
\bibitem{CHY5} F. Cachazo, S. He, and E.Y. Yuan,	JHEP {\bf 1507} (2015) 149 [arXiv:1412.3479].	
\bibitem{LY3}  C.S. Lam and Y-P. Yao,  Phys.~Rev.~D {\bf 93} (2016) 105008 [arXiv:1602.06419].
\bibitem{pw1} R. Huang, Y.-J. Du, B. Feng, JHEP {\bf 06} (2017) 133 [arXiv:1702.05840].
\bibitem{pw2} F. Teng, B. Feng, JHEP {\bf 05} (2017) 075 [arXiv:1703.01269].
\bibitem{pw3} Y.-J. Du, F. Teng, JHEP {\bf 04} (2017) 033 [arXiv:1703.05717]. 
\bibitem{pw4} S. He, O. Schlotterer, Y. Zhang, Nucl.~Phys. {\bf B03} (2018) 003 [arXiv:1706.00640].
\bibitem{pw5} X. Gao, S. He, Y. Zhang, JHEP {\bf 11} (2017) 144 [arXiv:1708.08701].
\bibitem{FG} F. Cachazo, H. Gomez,  JHEP {\bf 04} (2016) 108 [arXiv:1505.03571].
\bibitem{BBBD2} C. Baadsgaard, N. E. J. Bjerrum-Bohr, J. L. Bourjaily, and P. H. Damgaard,  
		JHEP {\bf 1509} (2015)136 [arXiv:1507.00997].
\bibitem{DG1} L. Dolan and P. Goddard, JHEP {\bf 1401} (2014) 152 [arXiv:1311.5200]
\bibitem{BBBD} C. Baadsgaard, N. E. J. Bjerrum-Bohr, J. L. Bourjaily, and P. H. Damgaard, JHEP {\bf 1509} (2015) 129 [arXiv:1506.06137].
\bibitem{LY2}  C.S. Lam and Y-P. Yao,  Phys.~Rev.~D {\bf 93} (2016) 105004 [arXiv:1512.05387].
\bibitem{zY} Y. Zhang, JHEP {\bf 07} (2017) 069 [arXiv:1610.05205].
\end{thebibliography}
\end{document}